\documentstyle[preprint,eqsecnum,aps,prb]{revtex}

\begin{document}
\title{Current-induced metallic behavior in Pr$_{0.5}$Ca$_{0.5}$MnO$_3$ thin films:
competition between Joule heating and nonlinear conduction mechanism.}
\author{P. Padhan$^1$, W. Prellier$^{1,\thanks{%
prellier@ismra.fr}}$, Ch.\ Simon$^1$, R.C. Budhani$^2$}
\address{$^1$Laboratoire CRISMAT, CNRS UMR 6508, 6 Bd du Mar\'{e}chal Juin,\\
F-14050 Caen Cedex, FRANCE.}
\address{$^2$Department of Physics, Indian Institute of Technology Kanpur,\\
Kanpur-208016, INDIA.}
\date{\today}
\maketitle

\begin{abstract}
Thin films of Pr$_{0.5}$Ca$_{0.5}$MnO$_3$ manganites exhibiting
charge/orbital-ordered properties with colossal magnetoresistance have been
synthesized by the pulsed laser deposition technique on both (100)-SrTiO$_3$
and (100)-LaAlO$_3$ substrates. The effects of current-induced
metallic-behavior of the films are investigated as a function of the
temperature and the magnetic field. Calculations based on a heat transfer
model across the substrate, and our resistivity measurements reveal effects
of Joule heating on charge transport over certain ranges of temperatures and
magnetic fields. Our results also indicate that a nonlinear conduction,
which cannot be explained by homogeneous Joule heating of the film, is
observed when the material is less resistive (%
\mbox{$<$}%
10$^{-2}$ $\Omega .$cm). The origin of this behavior is explained with a
model based on local thermal instabilities associated with phase-separation
mechanism and a change in the long range charge-ordered state.
\end{abstract}

\newpage

Mixed valence manganites exhibit several fascinating phenomena such as
colossal magnetoresistance (CMR), spin polarization, ordering of charge,
orbital and spin of Mn$^{3+}$ and Mn$^{4+}$ ions, and electronic phase
separation\cite{CMR1,CMR2}. Charge/orbital-ordering (CO/OO) phenomenon in
particular have been seen when the dopant concentration ($x$) is close to
the commensurate value x=0.5 in the reduced bandwidth systems. The
charge-ordering gap in these systems collapses upon application of external
perturbation like magnetic field, electric field, high pressure, optical
radiation and electron irradiation\cite{CO}. This results in a metal-like
transport below the charge-ordering transition temperature (T$_{CO}$). For
example, electric field leads to a nonlinear transport with hysteresis and
switching depending on current, temperature and magnetic field history.
These effects observed in different charge-ordered manganite systems, are
attributed to the formation of conducting ferromagnetic regions within the
antiferromagnetic medium\cite{1,2,3,4,5,6,7,8}. In fact, the transport in
these systems is often interpreted in terms of spin polarized tunneling
conduction across the insulating phase and a percolative conduction
mechanism. However, other explanations can be found in the literature, and
some of them are listed below. For example, Guha {\it et al.}\cite{2} have
attributed their observations of a non-linear transport in Nd$_{0.5}$Ca$%
_{0.5}$MnO$_3$ thin films to depinning of the charge/orbital-ordered state.
In contrast, Tokura {\it et al.}\cite{3} have suggested that the electric
field may directly influence the direction of the orbital ordering in the
highly insulating state and alter the magnetic state. Moreover Stankiewicz 
{\it et al.}\cite{4} have suggested the formation of a mixed state that is
less resistive and highly magnetized than the initial charge-order state.
Furthermore, Markovich {\it et al.}\cite{7} have suggested the condensate
effect of charge density waves or spin density waves to the increase in
nonlinear conduction observed in Pr$_{0.8}$Ca$_{0.2}$MnO$_3$.

In this article, we report the effects of electric current on the transport
properties of Pr$_{0.5}$Ca$_{0.5}$MnO$_3$ (PCMO) thin films. The objective
of this paper is to develop an understanding of the current-induced decrease
of the resistivity in these materials. Thus, we have performed detailed
measurements of resistivity as a function of current and temperature in the
presence or absence of a magnetic field. Our data reveal that Joule heating
effects and nonlinear conduction of a fundamental nature are observed
depending on temperature and the value of the resistivity. The origin of
this behavior is explained with a model based on a phase separated medium.

Thin films (\symbol{126}800\AA ) of PCMO are grown on (100)-oriented SrTiO$%
_3 $ (STO) and LaAlO$_3$ (LAO) substrates using pulsed laser deposition
technique. More details of deposition, x-ray diffraction and
magnetotransport studies on PCMO are described elsewhere \cite{8,9}. The
epitaxy and texture of these films were confirmed by X-ray diffraction using
Seifert XRD 3000P and Philips MRD X'pert. The four-probe resistivity ($\rho $%
) measurements were performed using nanovoltmeter (Keithley 182), current
source (Keithley 236 source measure unit) in the cryostat of a Physical
Properties Measurement System (PPMS) supplied by Quantum Design USA as a
function of the temperature ($T$) and magnetic field ($H$). For
magnetization measurements, we have used the Quantum Design SQUID
magnetometer. The sample was cooled to a desired temperature from room
temperature in the absence of electric and magnetic fields to perform
transport measurements. The magnetic field was applied parallel to the
substrate plane.

The pseudo cubic lattice parameter of bulk PCMO is $\symbol{126}$3.802\AA 
\cite{10}, which is larger than the lattice parameter of LAO (3.79\AA )\ and
smaller than the lattice parameter of STO (3.905\AA ). These lattice
parameters indicate that these PCMO films on LAO will be under in-plane
compression whereas on STO the in-plane stress will be expansive. Also, the
lattice mismatch of PCMO with STO (3.77\AA ) is larger as compared to that
with LAO (3.839\AA ). The different state of stress in the 800 \AA\ thick
films of PCMO on STO and LAO is not distinctive in the zero-field
temperature-dependent resistivity shown in Fig.1a and Fig.1b. For example,
the zero-field cooled (ZFC) resistivity of PCMO on STO below room
temperature shows thermally activated behavior down to \symbol{126}100 K
with an activation energy ($E_a$) of 0.29 eV, which is slightly larger than
its bulk value (0.2 eV \cite{11}). Similar values are found for LAO (0.31eV)
in agreement with the zero-field dependance of the resistivity reported in
Fig.1a and 1b. Note that upon cooling below 100 K, the resistance of the
samples (on STO and on LAO) is not measurable due to the limitation of the
voltmeter. The charge-ordering temperature (T$_{CO}$\symbol{126}270 K) is
characterized by a kink in resistivity and appears to be same for both types
of films. The opposite nature of stress in the films of PCMO on STO and LAO
is clearly observed in the magnetic-field dependance of the resistivity vs.
temperature. For example, the ZFC resistivity of the PCMO/STO sample at 9
tesla (Fig.1a) is thermally activated on cooling below room temperature down
to 120 K. A metallic-like behavior is observed in the temperature range of
120 K to 65 K and below 65 K, the thermally activated behavior reappears. As
previously stated, the situation (in the presence of an applied magnetic
field) is different for PCMO/LAO. The resistivity of this sample on cooling
in 7 tesla magnetic field below room temperature shows thermally activated
behavior down to \symbol{126}160 K with $E_a$\symbol{126}0.31 eV, and then
below 160 K the resistivity is metal-like down to 10 K. Similar major
differences have already been observed in these compositions of thin films
and explained by the difference in the orientation of the thin films with
respect to the substrate \cite{8,9}. Another difference between both
substrates is observed in magnetic measurements (Fig. 1c and Fig. 1d). The
magnetic moment (FC and ZFC) of PCMO on STO decreases slowly on heating from
10 K to 190 K. The magnetic moment then decreases slowly above 220 K up to
room temperature. A qualitatively similar behavior of the magnetization of
the PCMO films deposited on LAO is seen. Denoted by a kink in the
temperature-dependance of the magnetic moment, the N\'{e}el temperature ($%
T_N $) of the PCMO films on LAO and STO is 170 K and 190 K, respectively
(see Fig. 1). This is close to the value for bulk samples ($T_N$ \symbol{126}%
175 K) \cite{11}. These results indicate that there are some significant
differences in both the magnetotransport and magnetization data of PCMO\ on
STO\ and LAO\ {\bf (}Fig.1) as a result of the substrate-induced strains.

Though the transport measurements in thin films show melting of the
charge-ordered state at a lower magnetic field as compared to that of the
bulk (for details see \cite{8,9}), this is not an artifact of the Joule
heating as these measurements have been performed at a very low current (10 $%
\mu $A). Different approaches have been used in the literature to minimize
Joule heating in such measurements. Some of these are:

(i) Limit the width and thickness of the samples to a few microns. This
approach can give erroneous results as the characteristic correlation length
of the ordered/disordered state in CO manganites is quite large \cite{12,13}.

(ii) An external load resistor in series with the sample is used as a
current limiting device\cite{1,12,13,14}. This arrangement limits the
maximum electric field that can be applied across the sample with a power
supply of moderate voltage output (\symbol{126}100V).

(iii) A pulsed (\symbol{126}1 s) current technique has also been used to
reduce Joule heating. However, the pulsed mode is not suitable to study
hysteretic behavior. Since the electrical conduction in these systems
depends strongly on the current history, the hysteretic behavior may change
if the current is switched on and off during cycling. In order to avoid some
of these complications, we have first reduced the strength of the CO state
through application of a magnetic field and then measured the charge
transport at moderate electric field.

Figure 2(a) shows the temperature dependent resistivity $\rho (T)$ of a 
\symbol{126}800\AA\ thick PCMO film on LAO measured at 10 $\mu $A for
different values of the in-plane magnetic field. In the presence of a 5
tesla magnetic field, the resistivity of the sample upon cooling below room
temperature shows thermally activated behavior down to 120 K and then
becomes metal-like in the temperature window of 65 K to 120 K. Below 65 K
the resistivity is again thermally activated. A qualitatively similar
behavior with a wider metal-like window and lower magnitude of resistivity
is seen at higher magnetic fields (%
\mbox{$>$}%
5 tesla). Such features are typical of a CO-system where an external
magnetic field destroy the CO-state and induces a metallic behavior \cite{CO}
and have already been observed in thin films of PCMO \cite{8,9}.

A difficult problem one faces inevitably when measuring voltage-current
characteristics is the Joule heating of the samples\cite{15}. Thus, prior to
performing these current-voltage ($I-V$) measurements, we have estimated the
suitable limit for the electric field from a simple calculation of the
increase in the sample temperature due to Joule heating. The increase in
sample temperature ($\Delta T$) when the current $I$ flows through it was
estimated by taking into account the heat dissipation in the sample and heat
conduction by the substrate. Assuming the substrate is anchored to the
isothermal base of the PPMS, the increase in temperature can be approximated
as $\Delta T$ \symbol{126}$2P_l/$ $\kappa _{sub}$ where $P_l$ is the power
dissipated per unit length of the sample and $\kappa _{sub}$ is the thermal
conductivity of the substrate. The increase in temperature of the sample at
a constant thermal base temperature ($T$) can be expressed as $\Delta
T(T,I)=(2I^2\rho (T+\Delta T)/S\kappa _{sub}(T+\Delta T))$ where $S$ is the
cross section of the sample\cite{15}. The $\kappa _{sub}$ of LAO at 77 K and
197 K is $0.186$ $W$ $cm^{-1}K^{-1}$ and $0.143$ $W$ $cm^{-1}K^{-1}$,
respectively \cite{16}. Even though the thermal conductivity of LAO varies
with temperature, we have assumed an average value of $\kappa _{sub}$= $0.15$
$W$ $cm^{-1}K^{-1}$, in the temperature range of 10-170 K. Using the
experimental values $\rho (T)$ and $\kappa _{sub}$= $0.15$ $W$ $%
cm^{-1}K^{-1} $, the calculated temperature rise ($\Delta T$) at different
temperatures for the sample (0.5 x 0.1 x 8.10$^{-6}$) cm$^3$ on LAO at 5
tesla and 7 tesla is shown in GFigures 2b and 2c, respectively. The
corresponding increase in resistivity inferred from Fig. 1 (Panel a) is
shown in panels d and e, respectively. While $\Delta T(T,I)$ is quadratic in
current, its value remains small (%
\mbox{$<$}%
2K) even when $\sim $3000 $\mu $A current passes through the sample. The
corresponding changes in the resistivity are also small (less than 0.2 $%
\Omega .$cm under a magnetic field of 5T at 170\ K\ ; see Fig. 2d). Similar
calculations were performed for PCMO films on STO. In this case, however,
the calculated increase in temperature is much larger. This is due to the
high resistivity of PCMO on STO (10$^{-1}$ $\Omega .$cm) in the range of
100-150 K even under the application of a 9 tesla field (see Fig. 1a). For
this reason, we focus our study only on the sample grown on LAO.

To understand the collective effect of magnetic field and electric field on
transport properties, we have measured the resistivity of a PCMO film on LAO
at 100 K for different values of magnetic field (from 0 to 9 T) and various
driving currents (from 1 $\mu $A to 2000 $\mu $A). The magnetic field
dependence of resistivity at 100 K for 1, 10, 1000 and 2000 $\mu $A currents
is shown in Fig. 3(a). The resistivity measured at 1 $\mu $A first decreases
slowly as the magnetic field is increased from zero to a critical value $%
H^{*}$ $\sim $3.5 tesla, and then drops rapidly untill reaching the maximum
field of 7 tesla. As the magnetic field is reduced from 7 tesla, the
resistivity is irreversible up to the field $H_{irr}$ $\sim $1.5 tesla and
then becomes reversible. This hysteretic behavior is typical of a first
order transition as previously observed from CO\ compounds \cite{CO}. For
higher values of current, the zero-field and field dependent resistivity
shows the following four characteristics:

(i) below 150K, the resistivity in zero-field is lower for higher currents
(up to 1000 $\mu $A ; see Fig.3b)

(ii) the magnitude of resistivity is lower

(iii) $H_{irr}$ shifted to the higher fields

(iv) the MR at 2000 $\mu $A current is less than the MR with a current below
2000 $\mu $A.

The combine effect of electric field and magnetic field on ZFC resistivity
of PCMO as a function of temperature is shown in the Fig. 3(c and d). In
fact, three regimes are observed (I, II and III). At 5 tesla and 10 $\mu $A
current, the $\rho (T)$ below room temperature is thermally activated down
to 125 K (regime I), and in the temperature window of 65 K to 125 K, it has
metal-like behavior (regime II). Below 65 K, the $\rho (T)$ is increasing
upon cooling (regime III). However, a large drop in $\rho (T)$ is seen with
the increasing current in this temperature range (10-65K). A qualitatively
similar behavior of resistivity with the increasing current is seen in 7
tesla magnetic field (Fig.3d). Although the CO-phase in these systems phase
is destroyed by the magnetic field, the effect of electric current on the
resistivity at low temperatures is still prominent.

In order to address this non-linear transport, we have carried out
isothermal measurement of resistivity as a function of current. The
current-dependence of resistivity $\rho (I)$ of PCMO on LAO at 75 K in the
presence of 5, 6, 7 and 9 tesla is shown in the left hand panel of the Fig.
4. The resistivity at 5 tesla first decreases slowly up to 1000 $\mu $A and
then decreases much more rapidly. On decreasing the current below 2500 $\mu $%
A, the $\rho (I)$ goes through a peak at 950 $\mu $A where the resistivity
is higher than the value seen in the current increasing branch. On lowering
the current, the resistivity decreases and crosses the current increasing
branch at 300 $\mu $A. The behavior of $\rho (I)$ at 6 tesla is
significantly different. Here the resistivity goes through a minimum at a
critical current ($I_C\sim $1000 $\mu $A) and then increases rapidly with
the increase in current up to 2500 $\mu $A. The resistivity-current
decreasing of the branch $\rho (I)$ is lower than the forward branch over
the entire range of current. The hysteretic behavior of $\rho (I)$ at 7
tesla is qualitatively similar, with a higher $I_C$ ($\sim $1500 $\mu $A).
However, at 9 tesla the resistivity continues to drop with current without
any signatures of $I_C$ untill reaching the maximum current of 2500 $\mu $A.
The $\rho (I)$ of the same sample at 10, 50, 100 and 130 K and 7 tesla field
is shown in the right hand panel of Fig. 4. The hysteretic behavior of
resistivity as the current is swept from zero to 2500 $\mu $A and back to
zero, is seen at all temperatures with some subtle differences.

The observed behavior of resistivity of PCMO films as a function of
temperature, current and strength of magnetic field can be understood as
follows. The thermally activated behavior at the lowest temperature in the
ZFC $\rho $(T) at 7 tesla as seen in Fig.1(b) indicates the presence of a
charge-order gap. However, a long range CO-state does not form when we cool
the sample in 7 tesla field. While the magnetic field promotes the growth of
ferromagnetic clusters and reduces spin dependent intercluster scattering,
the electric field increases the ferromagnetic fraction as shown by the
increase of $H_{irr}$, $H^{*}$ and the reduction of resistivity with
increasing current as seen in Fig.3a.

These nonlinear behaviors of the resistivity with current suggest the
presence of other mechanisms. To understand this nonlinear transport in
PCMO, we have measured the $\rho $(I) in different regions of temperature in
the presence of a magnetic field. The isothermal $\rho $(I) shows hysteretic
behavior and it is symmetric with current provided the sample is cooled from
the room temperature to the desired temperature to measure both the
directions of current. In the paramagnetic and insulating state, $\rho $(I)
decreases slowly with temperature and is reversible with current. The
hysteretic behavior of $\rho $(I) at T%
\mbox{$<$}%
T$_C$ (Fig. 4) distinguished three zones in the $\rho $(T) curve, the
temperature zone close to the metal-to-insulator (M-I) transition,
metal-like $\rho $(T) zone, and low temperature insulating zone. These three
zones have been marked as region I, region II and region III respectively
and are shown in Fig. 3. In Fig.4, region (I) is observed at 130 K and 7 T
whereas region (III) corresponds to 10\ K\ and 50 K\ at 7 T. Region (II) is
seen at 75 K with a magnetic field from 5 to 9 T. On reducing current, the
resistivity stays relatively constant. Note that at zero current, the
resistivity is always lower than the initial one.

These data can be explained using the phase-separation scenario (PS)
scenario \cite{PS} which consists of a fine mixture of the two competing
ground states, the ferromagnetic (FM) metallic state and the charge/orbital
ordered insulator one\cite{PS,PS1,PS2,PS3}. The competition between the FM
metallic state due to the double-exchange and the antiferromagnetic (AFM)
nonmetallic state, due to the 1:1 ordering of Mn$^{3+}$/Mn$^{4+}$, has been
seen in RE$_{1-x}$Ca$_x$MnO$_3$ (RE=Pr, Nd)\cite{PS4}. Thus, we have assumed
that our PCMO film is a PS system. This is reasonable based both on our
previous experiments using electron spin resonance made on the same film
composition \cite{Sophie} and also on the magnetotransport measurements of
films which indicate that the metal-like regions are growing and that the
insulator-like regions are, simultaneously, shrinking in the presence of an
applied magnetic field (see Fig. 1a, 1b and 2a). If one compares the three
different regions (I), (II) and (III), the situation of the ground states is
different, due to the presence of magnetic field and the different
temperatures ranges\cite{CO} (this is observed in the value of resistivity
in Fig.3c and Fig.3d). For example, from the magnetic point of view, in
region (I), the film is in the PM state with a short range ordering \cite
{PS1} whereas in region (II), the AFM (COI) regions are present in the FM
(metal) matrix. In region (III), the AFM (COI) state is partially ordered.
The picture of the mechanism of current-induced is as follows.\ When current
is applied in the sample, it drives the electrons towards the lower
potential energy, which changes the effective electron-electron interaction
and modifies the charge distribution in the metal-like stripe regions that
stays when the current is reducing to 1 $\mu $A.\ This is evidenced in
Fig.5a where the resistivity shows an hysterestic behavior as a funtion of
the applied current. Depending on the values of the resistance, {\it i.e.},
the portion of metallic phase with respect to the insulating one, the
current will stabilize local thermal regions. When the current is reduced,
the sample is (magnetic) field cooled and remains in its small resistivity
value as one can see in Fig.1. This explains why the resistance decreases as
the current is reduced. Using this model, the current-dependance of the
resistivity in the three different regions can now be explained.

In region (I), the sample is mostly metallic, but the insulating regions are
growing due to the applied magnetic field and the value of temperature (130\
K which is just below the insulator-to-metal transition in Fig.2a). Below
1000 $\mu $A, the hysteretic behavior and the decrease of resistivity are
observed, indicating a melting of the CO\ state.\ However, the changes are
very small confirming that the CO\ state is almost completely melted with an
applied magnetic field of 7 T.\ Above 1000 $\mu $A, the reversal resistivity
drops are due to the homogeneous Joule heating.

In region (II), the resistivity increases as temperature increases (under
applied magnetic field).\ In addition, the metallic phase is growing as the
magnetic field increases, a signature of the melting of the CO\ state\cite
{CO}. When the current is below 1000 $\mu $A, most of the CO has been melted
although the value of magnetic field is not sufficent to melt it completely
(at least below 9T). Thus, the applied current will melt the regions that
were still CO. Since the fraction of these regions is very small , the
decrease is very small.\ On the contrary, with a magnetic field of 9 T, the
resisitivity is constant under applied current up to 500 $\mu $A confirming
that the CO\ is completely melted in accordance with previous reports\cite
{8,9}. Surprisingly above 1000 $\mu $A, the resistivity increases as the
current increases. This increase can be explained with thermal instabilities
because in this region, when the temperature increases, the resisitivity
increases as well.

In region (III), most of the sample is insulating because the magnetic field
is not sufficient to completely melt the CO state\cite{8,9}.\ When the
current is applied (up to 1000 $\mu $A), it also modifies the charge
distribution of the CO state, increasing the metallic regions\cite{Bulk1}.
However, the magnetic field will induce some metallic regions that, as seen
previously, will heat locally.

The situation of the $\rho $(I) at 75 K and 5 T is more complicated because
it is just at the boarder between the two regions (I) and (II). On
increasing the current, there is a decrease and then a jump at 1000 $\mu $A
due to Joule heating.\ In fact, the heating is so important, that the sample
is crossing the metal-to-insulator region (see Fig.2a).\ On decreasing the
current, it will follow that curve and this results in the peak in the $\rho 
$(I).

In fact, in the different regions the resistivity decreases as the current
increases up to critical value (I$_C$), and then either increases or
decreases rapidly (depending on the values of resistivity). The behavior can
thus be separated into two zones, below and above I$_C$ which can be
considered as a threshold. On Fig.5b, we have reported this critical current
which separates the two behaviors (thermal instabilities and non-linear
conduction). This limit is an image of the shape of the temperature and
magnetic field dependence of the resistivity.

We also have noted that the effect of current-induced decrease of
resistivity in manganite-based compounds is not a particular case of the
thin films and has already been studied in bulk samples\cite
{Bulk1,Bulk2,Bulk3}. However, in some cases, the values of current-inducing
a metallic behavior were much higher (on the order of mA or above) and the
real effects are probably obscured by Joule heating\cite{Bulk2,Bulk3}. In
other cases\cite{Bulk1}, the values of the current are lower compared to our
values (on the order of nA) which makest the comparison difficult. In this
latter case, they clearly observed a first order transition from a
charge-ordered state to a metallic state.

In conclusion, we have studied the effects of currents on the transport
properties of Pr$_{0.5}$Ca$_{0.5}$MnO$_3$ thin films grown on LaAlO$_3$and
SrTiO$_3$ substrates. We have performed detailed measurements of the
resistivity as a function of current and temperature in the presence or
absence of a magnetic field. Our data reveal that two mechanisms, local
heating and non linear conduction are observed depending on temperature and
the value of the resistivity. The origin of this behavior is explained with
a model based local thermal instabilities in the metallic percolation
regions resulting from the phase-separation system and a modification in the
long range charge-ordered state.

We greatly acknowledge financial support of Centre Franco-Indien pour la
Promotion de la Recherche Avancee/Indo-French Centre for the Promotion of
Advance Research (CEFIPRA/IFCPAR) under Project N${{}^{\circ }}$2808-1. We
also thank Dr. H. Eng, Dr. A. Wahl and Dr. A. Maignan for careful reading of
the article and helpful discussions during this work. 

\newpage Figures Captions:

Figure 1: (a) Zero-field-cooled temperature dependent resistivity of Pr$%
_{0.5}$Ca$_{0.5}$MnO$_3$ deposited on ($001$) oriented STO at $0$ and $9$
tesla magnetic field at $10\ \mu $A current. (b) Zero-field-cooled and
field-cooled temperature dependent resistivity of PCMO deposited on (001)
oriented LAO at $0$ and $7$ tesl$a$ magnetic field at $10$ $\mu $A current.
Panel (b) and (c) show the field-cooled magnetization of PCMO on STO and
LAO, respectively, at $1.5$ tesla magnetic field at various temperature.

Figure 2: Panel a shows the zero-field-cooled temperature dependent
resistivity of PCMO deposited on (001) oriented LAO at various magnetic
field with $10$ $\mu $A driving current. Lower panels (b) and (c) show the
estimated change in the resistivity while panel (d) and (e) show the
estimated change in the sample temperature due to the driving current in the
sample at different temperature at 5 tesla and 9 tesla magnetic field,
respectively.

Figure 3: Panel (a) shows the zero-field-cooled resistivity of the same
sample at various current as the magnetic field increases from zero to $7$
tesla and then decreases from $7$ $tesla$ to zero. Panel (b), (c) and (d)
show the zero-field-cooled temperature dependent resistivity of PCMO on LAO
at $0$, $5$ and $7$ $tesla$ magnetic field at $10$, $100$ and $1000$$\mu A$
driving current.

Figure 4: Left hand panel shows the zero-field-cooled resistivity of PCMO on
LAO at $75$$K$ at various magnetic field as the current increases from $1$ $%
\mu A$ to $2500$ $\mu $A followed by a decrease to $1$ $\mu $A. Right hand
panel show the zero-field-cooled resistivity of the same sample at $7$$tesla$%
magnetic field at $10$, $50$, $100$ and $130$ K respectively as the current
increases from $1$ $\mu $A to $2500$ $\mu $A and then decreases from $2500$ $%
\mu A$ to $1$ $\mu A$.

Figure 5: Panel (a) shows the zero-field-cooled resistivity of PCMO on LAO
at $7$ tesla magnetic field at $100$ K as the current increases from $1$$\mu
A$ to a different higher value followed by a decrease to $1$$\mu A$. Panel
(b) shows the critical current at $7$ tesla magnetic field at different
temperatures. The critical current at different magnetic field at $75$$K$ is
shown in the inset of panel (b). The critical current I$_C$ is the
excitation current at which the MR\ starts to be non-hysteretic in Fig.4.


\bigskip \newpage

\begin{references}
\bibitem{CMR1}  A.P.\ Ramirez, J. Phys.: Condens. Matter 9, 8171 (1997).

\bibitem{CMR2}  W.\ Prellier, Ph.\ Lecoeur and B.\ Mercey, J.\ Phys.\ Cond.\
Matter. 13, R915 (2001).

\bibitem{CO}  C.N.R. Rao, A. Arulraj, A. K. Cheetham, and B. Raveau, J.
Phys.: Condens. Matter 12, R83 (2000).

\bibitem{1}  A. Asamitsu, Y. Tomioka, H. Kuwahara and Y. Tokura, Nature 388,
50 (1997).

\bibitem{2}  Ayan Guha, Arindam Ghosh, A. K. Raychaudhuri, S. Parashar, A.
R. Raju, and C. N. R. Rao, Appl. Phys. Lett. 75, 3381 (1999).

\bibitem{3}  Y. Tokura and N. Nagaosa, Science 288, 462 (2000).

\bibitem{4}  J. Stankiewicz, J. Sese, J. Garcia, J. Blasco, and C. Rillo,
Phys. Rev. B 61, 11 236 (2000).

\bibitem{5}  S. Srivastava, N.K. Pandey, P. Padhan, and R.C. Budhani, Phys.
Rev. B 62, 13 868 (2000).

\bibitem{6}  S. Mercone, A. Wahl, Ch. Simon, and C. Martin, Phys. Rev. B 65,
214428 (2002).

\bibitem{7}  V. Markovich, I. Fita, A. I. Shames, R. Puzniak, E. Rozenberg,
C. Martin, A. Wisniewski, Y. Yuzhelevskii, A. Wahl, and G. Gorodetsky, Phys.
Rev. B 68, 094428 (2003).

\bibitem{8}  W. Prellier, A. M. Haghiri-Gosnet, B. Mercey, Ph. Lecoeur, M.
Hervieu, Ch. Simon, and B. Raveau, Appl. Phys. Lett. 77, 1023 (2000).

\bibitem{9}  W. Prellier, Ch. Simon, A. M. Haghiri-Gosnet, B. Mercey, and B.
Raveau, Phys. Rev. B 62, R16337 (2000).

\bibitem{10}  Z. Jirak, S. Krupicka, Z. Simsa, M. Doulka, and S. Vratislma,
J. Magn. Magn. Mater. 53, 153 (1985).

\bibitem{11}  N. Lavrov, I. Tsukada, and Yoichi Ando, Phys. Rev. B 68,
094506 (2003).

\bibitem{12}  M. Fiebig, K. Miyano, Y. Tomioka, and Y. Tokura, Science 280,
1925 (1998).

\bibitem{13}  S. Yamanouchi, Y. Taguchi, and Y. Tokura, Phys. Rev. Lett. 83,
5555 (1999).

\bibitem{14}  K. Hatsuda, T. Kimura, and Y. Tokura, Appl. Phys. Lett. 83,
3329 (2003).

\bibitem{15}  A.N.\ Lavrov, I.\ Tsukuda and Y.\ Ando, Phys.\ Rev.\ B 68,
094506 (2003).

\bibitem{16}  Peter C. Michael, John U. Trefny, and Baki Yarar, J. Appl.
Phys. 72, 107 (1992).

\bibitem{PS}  for a review see: A.\ Moreo, S.\ Yunoki, E.\ Dagotto, Science
283, 2034 (1999).

\bibitem{PS1}  M. Uehara, S. Mori, C.H. Chen and S.-W. Cheong. Nature 399,
560 (1999); S. Mori, T. Asaka, and Y. Matsui, J. Electron Microsc. (Tokyo)
51, 225, (2002).

\bibitem{PS2}  L. Zhang, C. Israel, A. Biswas, R. L. Greene, and A. Lozanne,
Science, 298, 805 (2002).

\bibitem{PS3}  M.\ F\"{a}th, S.\ Freisem, A.A.\ Menovsky, Y.\ Tomioka, J.\
Aarts, J.A.\ Mydosh, Science 285, 1540 (1999).

\bibitem{PS4}  T.\ Tomioka, A.\ Asamitsu, H.\ Kuwahara, Y.\ Moritomo, M.\
Kasai, R.\ Kumai, Y.\ Tokura, Physica B 237-238 (1997) 6.

\bibitem{Sophie}  S.\ de Brion, G.\ Storch, G.\ Chouteau, A.\ Janossy, W.\
Prellier and E.\ Rauwel-Buzin, Eur.\ Phys.\ J.\ B\ 33, 413 (2003).

\bibitem{Bulk1}  A. Asamitsu, Y. Moritomo, Y. Tomioka, T. Arima, and Y.
Tokura, Nature 388, 50 (1997).

\bibitem{Bulk2}  S.\ Parashar, L. Sudheendra, A.R.\ Raju and C.N.R.\ Rao,
J.\ Appl.\ Phys. 95, 2181 (2004).

\bibitem{Bulk3}  L. Sudheendra and C.N.R.\ Rao, J.\ Appl.\ Phys. 95, 2767
(2003).
\end{references}
\end{document}